\documentclass[twoside,slac_one]{revtex4}
\usepackage{graphicx}
\usepackage{fancyhdr}
\usepackage{amsmath} 
\usepackage{bm}
\usepackage{amsxtra}
\usepackage{amssymb}
\usepackage{amsthm}
\usepackage{latexsym}
\usepackage{lscape}
\usepackage{natbib}

\usepackage{dcolumn}   

\def\slashchar#1{\setbox0=\hbox{$#1$}           
  \dimen0=\wd0                                 
  \setbox1=\hbox{/} \dimen1=\wd1               
  \ifdim\dimen0>\dimen1                        
     \rlap{\hbox to \dimen0{\hfil/\hfil}}      
     #1                                        
  \else                                        
     \rlap{\hbox to \dimen1{\hfil$#1$\hfil}}   
     /                                         
  \fi}                                         %
\def\met{\slashchar{E}_T}
\def\slashchar#1{\setbox0=\hbox{$#1$}           
  \dimen0=\wd0                                 
  \setbox1=\hbox{/} \dimen1=\wd1               
  \ifdim\dimen0>\dimen1                        
     \rlap{\hbox to \dimen0{\hfil/\hfil}}      
     #1                                        
  \else                                        
     \rlap{\hbox to \dimen1{\hfil$#1$\hfil}}   
     /                                         
  \fi}                                         %
\def\mpx{\slashchar{p}_x}                    %
\def\slashchar#1{\setbox0=\hbox{$#1$}           
  \dimen0=\wd0                                 
  \setbox1=\hbox{/} \dimen1=\wd1               
  \ifdim\dimen0>\dimen1                        
     \rlap{\hbox to \dimen0{\hfil/\hfil}}      
     #1                                        
  \else                                        
     \rlap{\hbox to \dimen1{\hfil$#1$\hfil}}   
     /                                         
  \fi}                                         %
\def\mpy{\slashchar{p}_y}                    %

\newcommand{\MLQ}{\ensuremath{M_{LQ}}}
\newcommand{\mlq}{\ensuremath{M_{LQ}}}

\begin{document}

\title{Searches for vector-like quarks and leptoquarks at D0}

%

\author{L.~Zivkovic}
\affiliation{Department of Physics and Astronomy, Brown University, Providence, RI, USA}

\begin{abstract}
We report on a search for vector-like quarks and leptoquarks at D0. In the absence of any significant excess over the expectations, 
we present the most stringent limits to date.
\end{abstract}

\maketitle

\thispagestyle{fancy}


\section{Introduction}

The standard model (SM) of particle physics accurately describes interactions at the electroweak scale. Several theories are proposed to describe physics 
beyond the SM.
Every new theory introduces a new spectrum of particles.
The D0 experiment at the Fermilab Tevatron Collider has searched for many of these new particles. We present here a search for first generation leptoquarks and 
vector-like quarks.


\section{Search for First Generation Leptoquarks}
\subsection{Motivation}
Leptoquarks ($LQ$s)~\cite{Acosta:1999ws,Nakamura:2010zzi} are predicted by 
many extensions of SM, such as supersymmetry~\cite{Martin:1997ns}, grand unified theories~\cite{1974PhRvL-32-438G},
 and string theory~\cite{Hewet:1989}. $LQ$s
are mediating bosons that allow leptons and quarks to interact with each other. Although $LQ$s can be scalar or vector fields, this proceedings will focus on 
scalar particles. At the Fermilab Tevatron Collider $LQ$s are produced as leptoquark-antileptoquark
pairs via quark-antiquark annihilation and gluon-gluon fusion. The production cross section is known at next-to-leading order
(NLO)~\cite{Kramer:1997hh}. In the low energy limit there 
is no intergenerational mixing, thus we search for the first generation $LQ$ pair production that further
decays to a pair of the first generation lepton and quark. In this paper, we report result where one $LQ$
decays to $eq$ and the other to $\nu_e q'$ (charge
conjugate states are assumed in the paper)~\cite{Abazov:2011qj}. 
If $\beta={\cal BR} (LQ\to eq)$ then $\sigma\times{\cal BR} (LQLQ\to eq\nu_e q')$ is maximized for $\beta=0.5$.

Limits on the production of first generation leptoquarks
have been reported by the DELPHI~\cite{Abreu:1998fw}, OPAL~\cite{Abbiendi:2003iv,Abbiendi:2001sw}, H1~\cite{Aktas:2005pr,Collaboration:2011qa}, ZEUS~\cite{Chekanov:2003af}, CDF~\cite{Acosta:2005ge}, and D0~\cite{Abazov:2009gf} Collaborations. 
Recently, CMS~\cite{Khachatryan:2010mp,Chatrchyan:2011ar}, and 
ATLAS~\cite{Collaboration:2011uv} published the first searches for scalar $LQ$ pair production at the CERN LHC.

\subsection{Analysis}

The D0 detector is described elsewhere~\cite{Abazov:2005pn,Abolins:2007yz,Angstadt:2009ie}. 
In this proceedings we report the result from 5.4~fb$^{-1}$ of data collected between 2002 and 2009. Signal and SM background 
processes that contain real electrons are modeled with Monte Carlo (MC) and include $V$+jets ($V=W,Z$), $t\bar{t}$, single top and diboson 
($WW,WZ,ZZ$) processes. 
Diboson processes are generated with {\sc pythia}~\cite{Sjostrand:2006za}, 
and their cross sections are calculated at next-to-leading order (NLO). $V$+jets and $t\bar{t}$ are produced with {\sc alpgen}~\cite{Mangano:2002ea},
interfaced to {\sc pythia} for subsequent parton showering and hadronization and their cross sections are known at next-to-next-to-leading order.
Multijet (MJ) background, where a jet mimics an electron, is estimated from data~\cite{Abazov:2008kt}.
Scalar leptoquark pair MC samples are
generated using {\sc pythia}  for
different $LQ$ masses between 200 and 360 GeV. The corresponding
cross sections at NLO are listed in Table~\ref{tab:lqxs}.

\begin{table*}[ht]
\begin{center}
\caption{Scalar $LQ$ pair production cross sections, calculated at NLO, for different $\MLQ$~\cite{Kramer:1997hh}.}
  \begin{tabular}{ | c | c|c|c|c|c|c|c|c|c|c|c|c|c|c|c | }
  \hline
$\mlq$ (GeV) & 200 & 210 & 220 & 230 & 240 & 250 & 260 & 270 & 280 & 290 & 300 & 310 & 320 & 340 & 360 \\
\hline
 $\sigma$ (fb) & 268 & 193 & 141 & 103 & 76 & 56 & 42 & 31 & 23 & 17 & 13 & 10 & 7.4 & 4.2 & 2.4\\
\hline 
  \end{tabular}
\label{tab:lqxs}
\end{center}
\end{table*}

We explored ways to correctly pair jets and $e$ or $\nu_e$ which are originating from the same $LQ$. We did not put any requirement on the number of jets,
 but we considered the two leading in $p_T$ jets for pairing. Thus, there are two possible combinations, pairing the leading jet to either 
 the $e$ or the $\nu_e$.
 We tried four different methods:
 \begin{itemize}
 \item matching by minimizing differences in $p_T$ from the combination of (jet,$e$) and (jet,$\nu_e$);
 \item reconstructing the $LQ$ from the both combinations, and pick the combination such that the distance in transverse plane, $\Delta\phi(LQ_1,LQ_2)$,
  is closest to $\pi$;
 \item matching by minimizing $\Delta\phi$ between the decay products of the $LQ$s;
 \item matching by minimizing the differences in transverse mass, $m_T$, reconstructed 
 from (jet,$e$) and (jet,$\nu_e$), since the $LQ$s are produced with the same mass.
 \end{itemize}
 The most effective algorithm, with success rate of $\sim 75\%$ is the one requiring that the differences between transverse mass are minimal.
 
\begin{figure*}[ht]
\centering
\includegraphics[width=80mm]{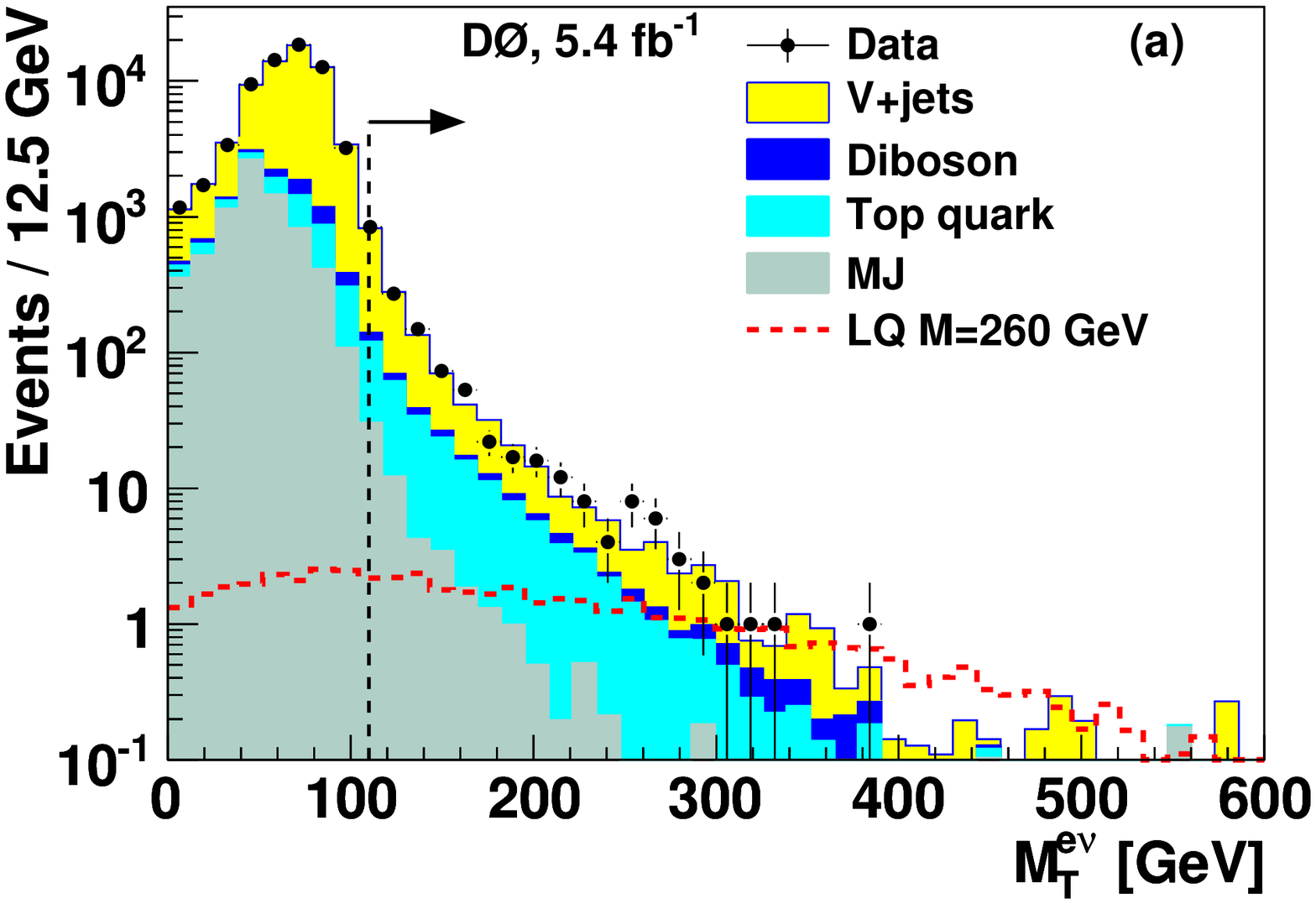}
\includegraphics[width=80mm]{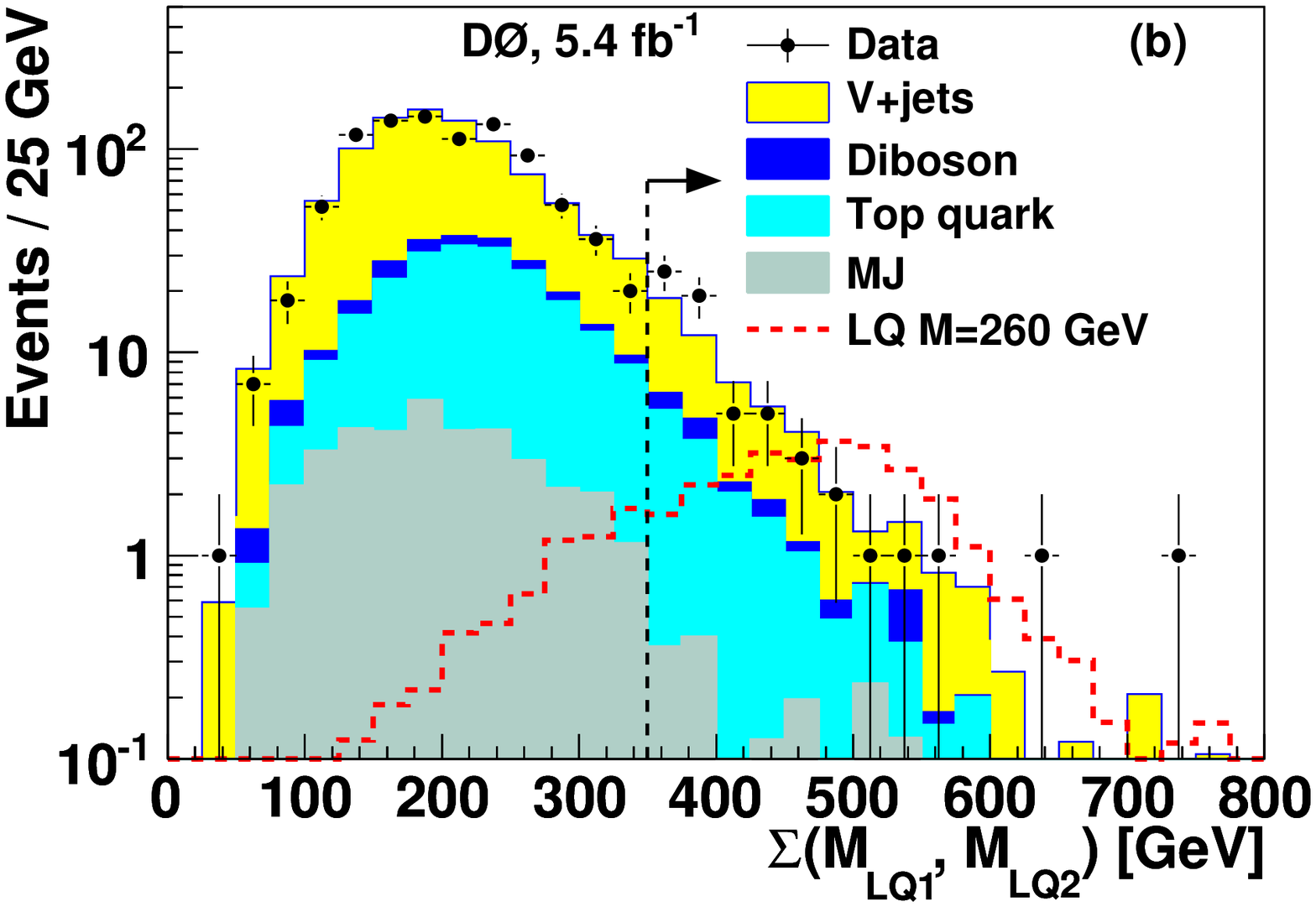}
\includegraphics[width=80mm]{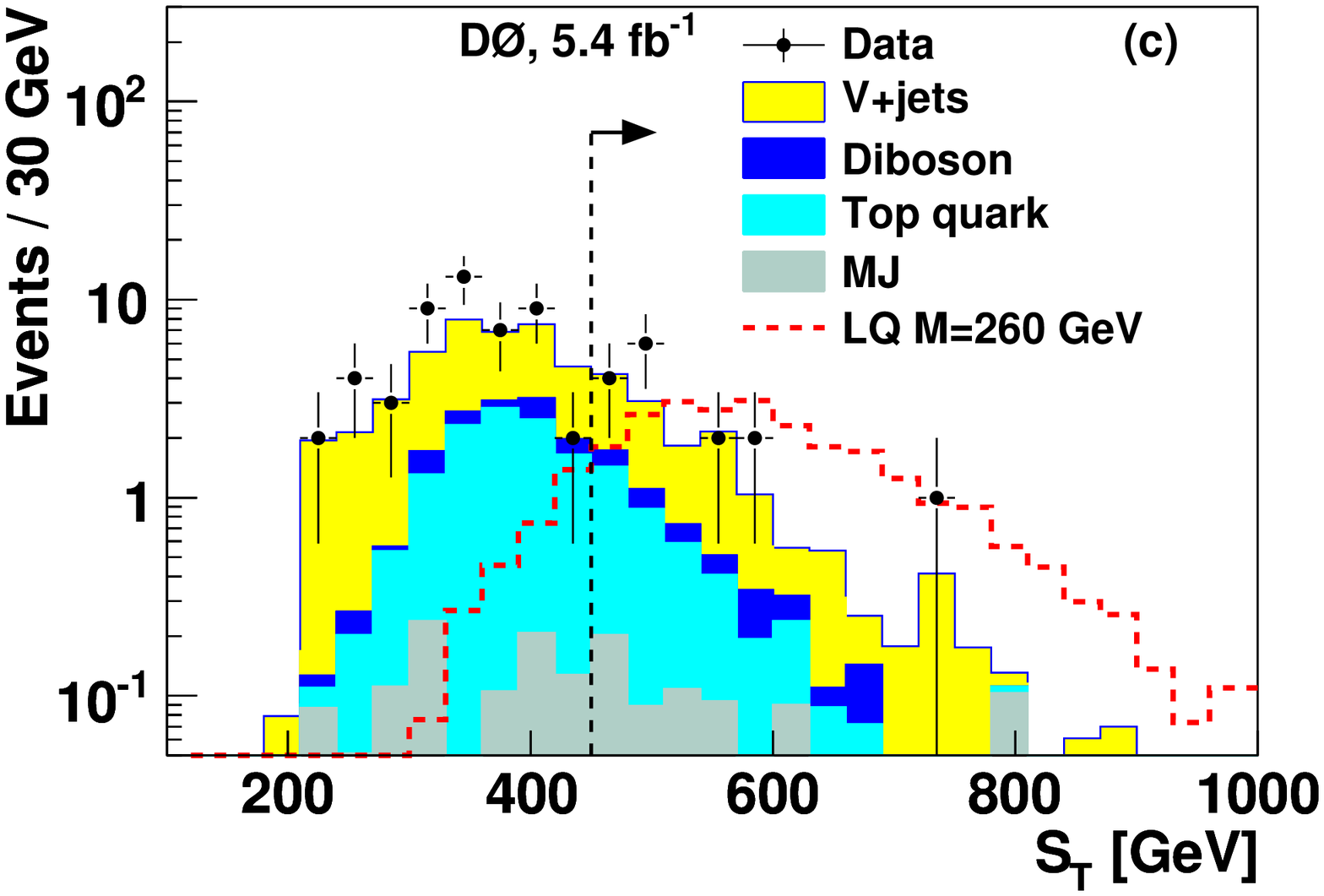}
\caption{ (a) $M_{T}^{e\nu_e}$ distribution after preselection,
  (b) $\sum{\mlq}$ for $M_{T}^{e\nu_e}>110$~GeV, (c)
  the $S_T$ for $M_{T}^{e\nu_e}>110$~GeV and $\sum{\mlq}>350$~GeV,
 which is used to set an upper limit on the $LQ$ pair production
  cross section after the final selection.} \label{fig:lq1}
\end{figure*}

 Events are selected if they have one electron with $p_T>15$~GeV, missing transverse energy $\met>15$~GeV and at least two jets with $p_T>20$~GeV. To suppress MJ background,
 it is further required that $\met/50 + M_{T}^{e\nu_e}/70 \geq 1$, where $M_{T}^{e\nu_e}$ is the transverse mass of the $(e,\nu_e)$ combination, 
 and $\met$ and $M_{T}^{e\nu_e}$ are in GeV.
 After these requirements, we observe 65992 data events, while we expect
$65703\pm 61(\rm stat)\pm 5958(sys)$ from SM background and
$50.4\pm 0.4(\rm stat)\pm 6.8(sys)$ events from scalar $LQ$
production for $\MLQ=260$~GeV and
$\beta=0.5$. Figure~\ref{fig:lq1}(a) shows the $M_{T}^{e\nu_e}$ distribution for the data and SM
processes.
To suppress the dominant background at this stage, $V$+jets, we select events that fulfill $M_{T}^{e\nu_e} \geq 110$ GeV.
We use the pairing algorithm described previously to reconstruct $\MLQ$. Since the $z$ component of the neutrino momentum is not measurable, for the 
$LQ\rightarrow \nu_e q'$ we reconstruct the visible part of the $\mlq=M(\text{jet}+\nu_{\text{vis}})$,
 where the four vector of
$\nu_{\text{vis}}$ is given as $(\mpx, \mpy, 0,
\met)$. Figure~\ref{fig:lq1}(b) shows the distribution of the sum $\sum{\mlq}$ of the invariant mass
of the decay $LQ\rightarrow e q$ and the visible mass of the
decay $LQ\rightarrow \nu_e q'$ after the requirement $M_{T}^{e\nu_e}\geq 110$~GeV. We then use
$\sum{\mlq}$ to reduce SM backgrounds, further requiring that $\sum{\mlq}>350$~GeV.
The final requirement is imposed on the scalar sum of the $p_T$ of lepton, neutrino and two jets, $S_T$, shown in Fig.~\ref{fig:lq1}(c) after the previous 
selection. We select events with $S_T>450$~GeV.  Event count after each selection requirement is shown in Table~\ref{tab:event_count_after_cuts}.

\begin{table}[ht]
\begin{center}
 \caption{Event counts and the predicted number of signal events for $\MLQ=260$~GeV and $\beta=0.5$ after each selection requirement.}
 \begin{tabular}{|l|c|c|c|}
\hline
 & Data  & Total background &  Signal \\
  \hline

  Preselection        & 65992 & $65703\pm 5958$  &  $50 \pm  7$ \\
 $M_{T}^{e\nu_e} > 110$ GeV & 990 & $986 \pm 82$  &  $34 \pm 5$  \\
  $\sum{\MLQ} > 350 $ GeV & 64 & $55 \pm 4$  &  $27 \pm 4$  \\
  $S_{T} > 450 $ GeV         & 15 & $15 \pm  1$  &   $24 \pm 3$  \\
\hline

  \end{tabular}
  \label{tab:event_count_after_cuts}
\end{center}
\end{table}

\subsection{Result}
As shown in Fig.~\ref{fig:lq1}(c) after the final requirement, we don't observe any significant excess, so we proceed to set limits. 
For each generated $\MLQ$, the limit is calculated at the 95\% C.L. using the
semi-frequentist $CL_s$ method based on a Poisson log-likelihood test
statistic~\cite{Junk:1999kv}.
Signal and background normalizations
and shape variations due to systematic uncertainties are incorporated
assuming Gaussian priors.

\begin{figure}[ht]
\centering
\includegraphics[width=80mm]{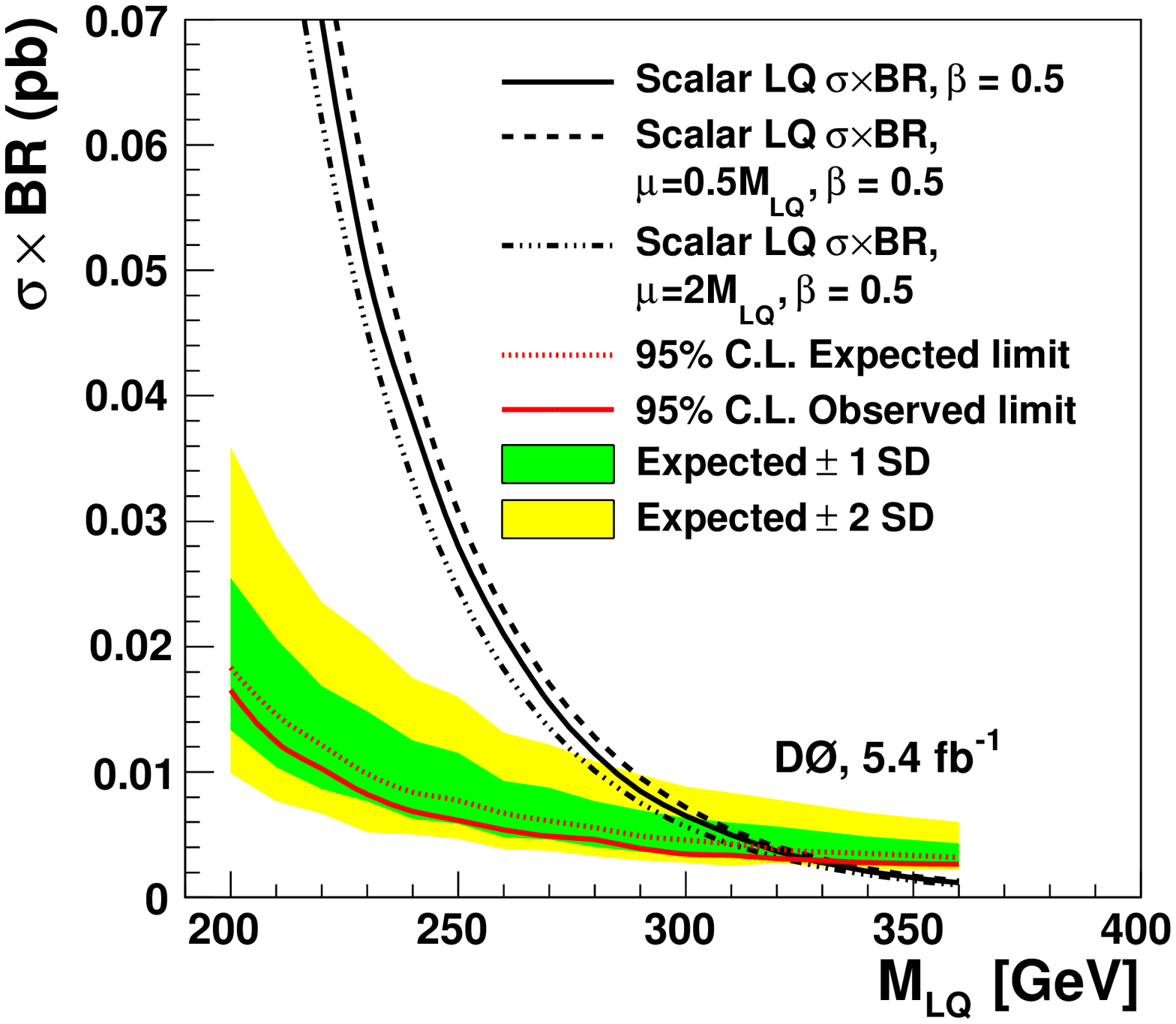}
\caption{Expected and observed upper limits calculated at the $95\%$
  C.L. on the $LQ$ cross section as a function of $\MLQ$ for a scalar
  $LQ$ compared with the NLO prediction for $\beta=0.5$. The NLO
  cross section is shown for different choices of the renormalization
  and factorization scales, $\mu=M_{LQ}$, $\mu=0.5\times M_{LQ}$, and $\mu=2\times M_{LQ}$.} \label{fig:lim1d}
\end{figure}

\begin{figure}[ht]
\centering
\includegraphics[width=80mm]{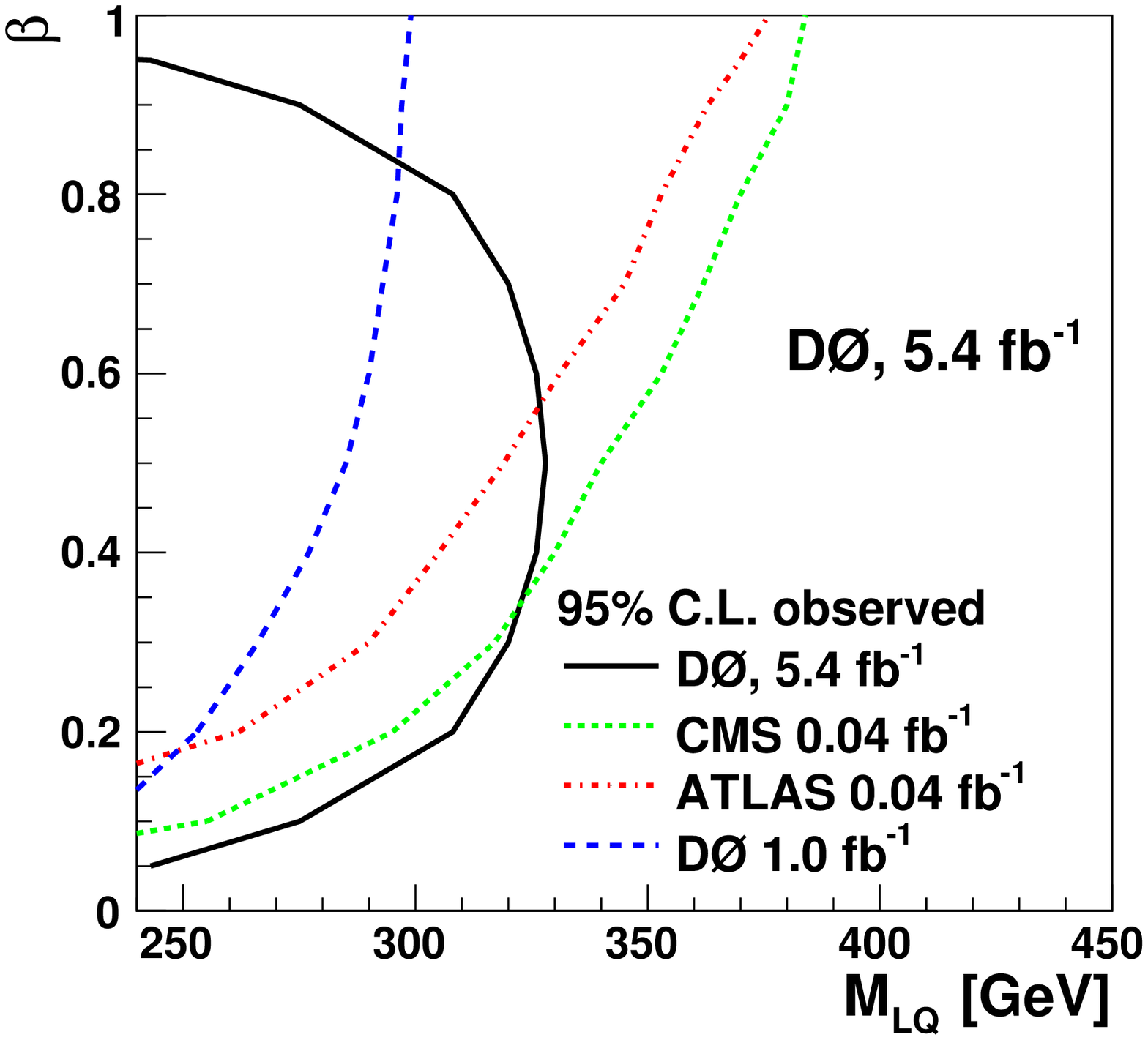}
\caption{ $95\%$ C.L. observed limit  for $\mu=M_{LQ}$
on the $LQ$ mass as a function of $\beta$ compared
  with the previous D0 result~\cite{Abazov:2009gf}, and CMS~\cite{Khachatryan:2010mp,Chatrchyan:2011ar} and
   ATLAS~\cite{Collaboration:2011uv} results.} \label{fig:lim2d}
\end{figure}

Figure~\ref{fig:lim1d} shows the limits on  the cross section multiplied by the branching fraction and
the theoretical $LQ$ cross section for $\beta=0.5$.
We exclude the production of first generation $LQ$s with
$\MLQ<326$~GeV for $\beta=0.5$ at the $95\%$ C.L.
We also determined the limit as a function of $\beta$, which is shown in Fig.~\ref{fig:lim2d} and compared with previous D0~\cite{Abazov:2009gf},
and recent ATLAS~\cite{Collaboration:2011uv} and CMS~\cite{Khachatryan:2010mp,Chatrchyan:2011ar} results.

\section{Search for Vector-Like Quarks}
\subsection{Motivation}
Many new theories predict the existence of
vector-like quarks, massive particles which share
many characteristics with SM quarks. They include, among others, little Higgs models~\cite{ArkaniHamed:2002qy}, 
warped extra dimensions~\cite{Randall:1999ee},
and universal extra dimensions~\cite{Appelquist:2000nn}.
Vector-like quarks are fermions (despite the name) and their left-and right-handed components transform in the same way under 
$SU(3)\times SU(2)_L\times U(1)$.
In $p\bar{p}$ collisions such as at the Tevatron Collider vector-like quarks can be produced in pairs via the strong interaction, 
or singly via the electroweak interaction.
In some scenarios (e.g. warped extra
dimensions), corrections to SM quark
couplings due to mixing with vector-like
quarks can cancel, and then single electroweak production at Tevatron will be enhanced~\cite{Atre:2008iu}.
Electroweak couplings between vector-like quarks and SM quarks depend on a parameter $\kappa_{qQ}$:
\begin{equation}
\kappa_{qQ}=\frac{v}{m_Q}\tilde{\kappa}_{qQ}
\end{equation}
where $v$ is the vacuum expectation value of the SM Higgs
field, $m_Q$ is the mass of the vector-like quark, and $\tilde{\kappa}_{qQ}$
is the coupling strength.
We present here results from a search of a singly produced vector-like quark in 5.4~fb$^{-1}$ of data collected with D0 detector~\cite{Abazov:2010ku}.

\subsection{Analysis}
Vector-like quarks can either decay to $W+q$ or $Z+q$, and they are always produced together with another quark.
Thus, we are looking at final states consistent with two jets and either $W\to l\nu$, i.e. single lepton channel, or $Z\to ll$, i.e. dilepton channel.
The main background for the single lepton channel is $W$+jets, while $Z$+jets is the main background for the dilepton channel. Other backgrounds 
include $t\bar{t}$, single top, diboson, and MJ. $V$+jets and $t\bar{t}$ are modeled with {\sc alpgen} interfaced with {\sc pythia}, 
single top with {\sc comphep}~\cite{Boos:2004kh}, 
and diboson with {\sc pythia}. MJ backgrounds are estimated using data driven techniques.
Signal samples are generated using {\sc madgraph}~\cite{Alwall:2007st},
with CTEQ6L1~\cite{Pumplin:2002vw} parton distribution functions, LO
cross sections from Ref.~\cite{Atre:2008iu} and the vector-like quark resonance
widths calculated with {\sc bridge}~\cite{Meade:2007js}. Subsequent parton
shower evolution is generated with {\sc pythia}.
For simplicity, we assume $\tilde{\kappa}_{uD}=1$, $\tilde{\kappa}_{uU}=\sqrt{2}$ and $\tilde{\kappa}_{dU}=\tilde{\kappa}_{dD}=0$,
i.e. ${\cal BR}(QD\to Wq) ={\cal BR}(QU\to Zq)= 100\%$, where $QU$ and $QD$ are up-type and down-type vector-like quark, respectively.

In the dilepton channel events are selected if they have two oppositely charged leptons with $p_T>15$~GeV, and if dilepton invariant mass is consistent with 
the mass of the $Z$ boson, i.e. $70<M_{ll}<110$~GeV. It is further required that events contain at least two jets with $p_T>20$~GeV and no significant $\met$,
i.e. $\met<50$~GeV. Since the signal in this final state is a heavy resonance decaying to a $Z$ boson and a jet, we further optimize cuts by requiring that 
$p_T^{ll}>100$~GeV, transverse momentum of the leading jet $p_T>100$~GeV, and that distance between the two leptons $\Delta R(l,l)<2.0$, where 
$\Delta R=\sqrt{\Delta\phi^2+\Delta\eta^2}$.

In the single lepton channel we select events with one lepton with $p_T>15$~GeV, $\met>15$~GeV and at least two jets with $p_T>20$~GeV. To suppress
MJ background we further require $2\times M_T^W+\met > 80$ ($M_T^W$ and $\met$ are in GeV). Selection requirements are optimized in the single lepton 
channel requiring that lepton $p_T>50$~GeV, leading jet $p_T>100$~GeV, $\met>40(50)$~GeV for electron(muon) channel, $M_T^W<150$~GeV and 
$\Delta\phi(l,\met)<2.0$. In $Qq\to Wqq$ events the second jet originates from SM quark produced in association with a vector-like
 quark, thus it will be forward and 
relatively soft. The direction of this jet is strongly correlated with
the charge of the produced vector-like quark, and thus
also with the charge of the lepton from its decay. The final requirement is then $Q_l\times\eta_{jet2}>0$, where $Q_l$ is the lepton charge and
$\eta_{jet2}$ is the pseudorapidity of the second jet. The efficiency of this cut is $\sim 85\%$ for signal events and $\sim 50\%$ for SM backgrounds.

\subsection{Results}
\begin{figure*}[ht]
\centering
\includegraphics[width=80mm]{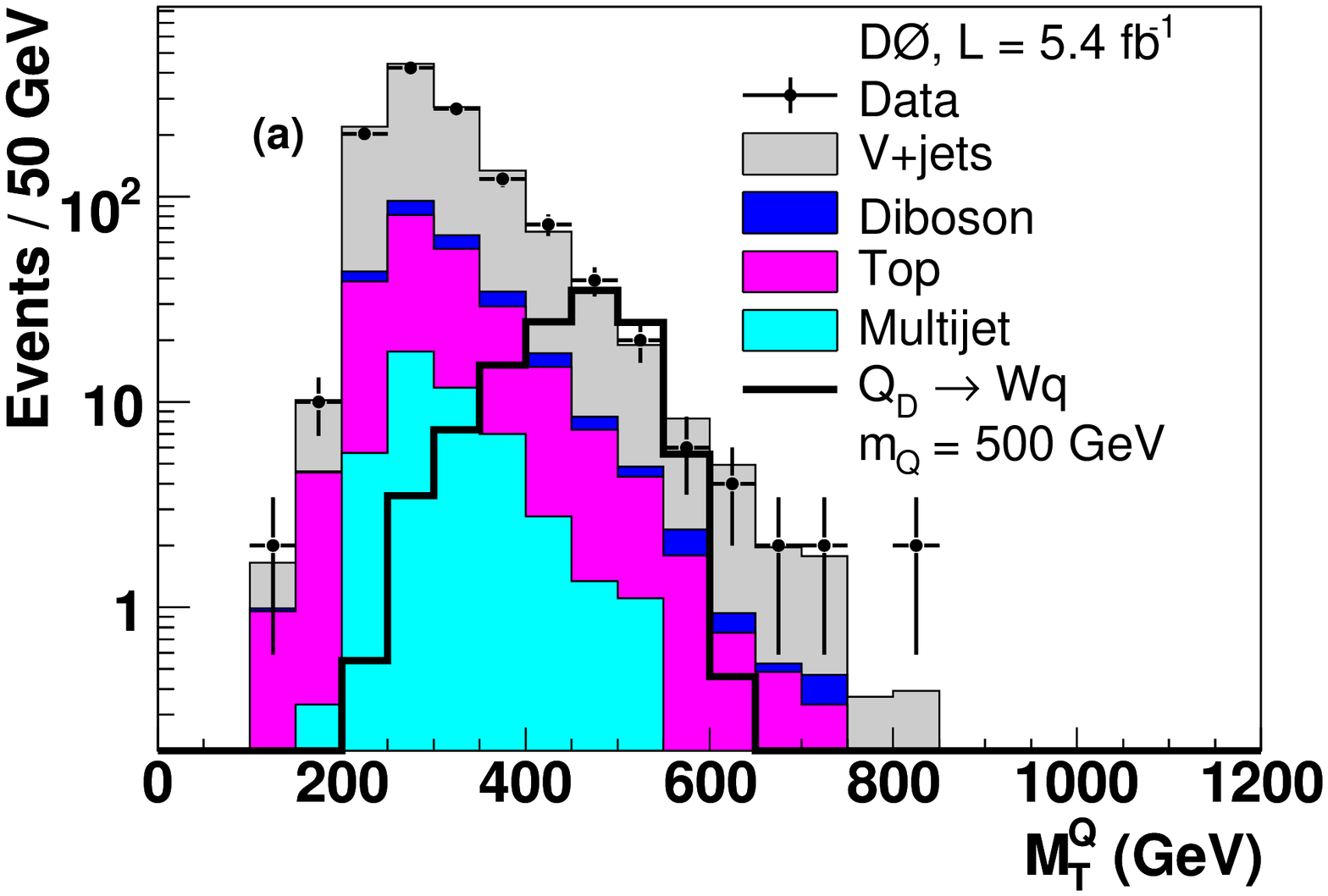}
\includegraphics[width=80mm]{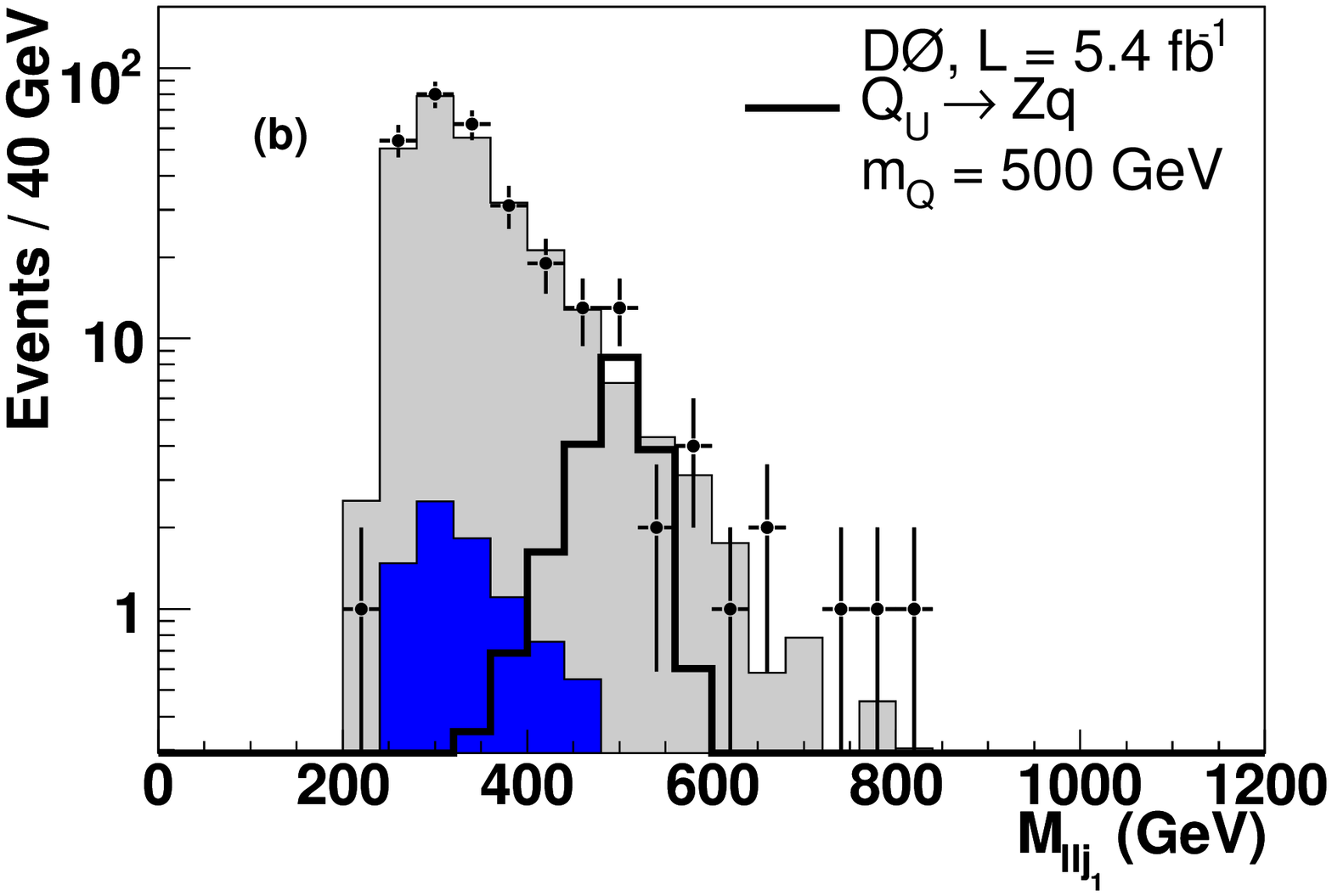}
\caption{(a) Vector-like quark transverse mass and (b) vector-like quark mass for the single lepton and dilepton channels, respectively.} \label{fig:vq1}
\end{figure*}

\begin{figure}[ht]
\centering
\includegraphics[width=80mm]{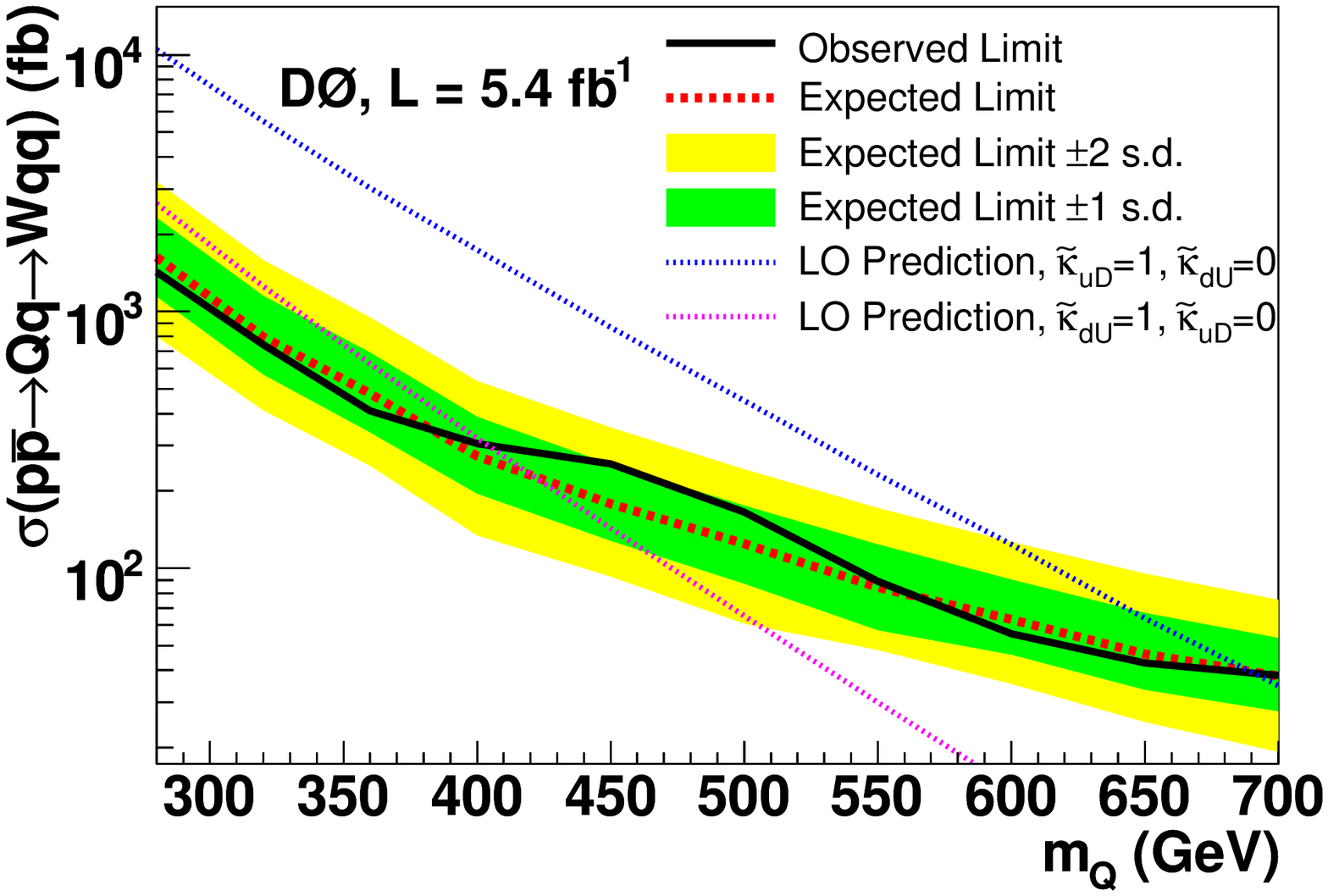}
\caption{Limit on the production cross section for a vector-like quark $Q\to Wq$ as a function of $m_Q$, compared to LO predictions of vector-like quark production with different $\tilde{\kappa}_qQ$.} \label{fig:vq2}
\end{figure}

\begin{figure}[ht]
\centering
\includegraphics[width=80mm]{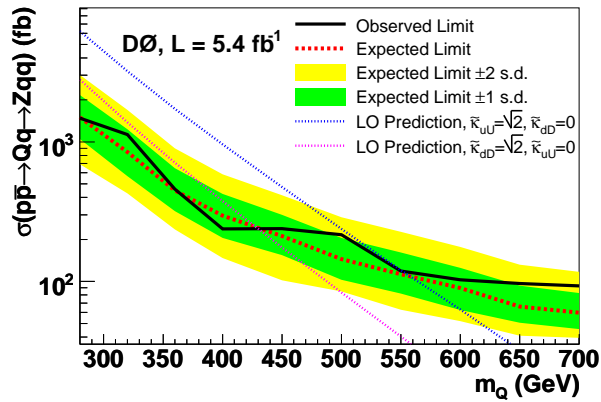}
\caption{Limit on the production cross section for a vector-like quark $Q\to Zq$ as a function of $m_Q$, compared to LO predictions of vector-like quark production with different $\tilde{\kappa}_qQ$} \label{fig:vq3}
\end{figure}

Figure~\ref{fig:vq1}(a) shows the reconstructed transverse mass of a vector-like quark, 
i.e. transverse mass of the lepton, $\met$ and leading jet, for the single lepton 
channel, and Fig.~\ref{fig:vq1}(b) shows the reconstructed invariant mass of a vector-like quark, i.e. invariant mass of the two leptons and leading jet, for the 
dilepton channel. Since we do not observe any significant excess in data over SM backgrounds, we proceed to set 95\% C.L. limits on the production cross 
section for a single vector-like heavy quarks. We employ semi-frequentist $CL_s$ method based on a Poisson log-likelihood test
statistic~\cite{Junk:1999kv}.
Figure~\ref{fig:vq2} and \ref{fig:vq3} show 95\% C.L. limits on the vector-like quark
production cross sections for the single lepton and dilepton channels, respectively.
We exclude at 95\% C.L. vector-like heavy quarks with masses below 693 GeV for $Q\to Wq$ and with masses below 449 for $Q\to Zq$,
 assuming $\kappa_{qQ}=1$.

\section{Summary}
In summary, we present results from the search for first generation leptoquarks and heavy vector-like quarks.
The observed data is consistent with the expectation from SM backgrounds.
We exclude pair production of scalar leptoquarks with masses below 326 GeV for $\beta=0.5$. We set the most stringent limits on pair production of scalar 
leptoquarks for $\beta<0.3$.
We also exclude vector-like heavy quarks with masses below 693 GeV for $Q\to Wq$ and with masses below 449 for $Q\to Zq$, respectively. 
These limits are the best to date.


\begin{acknowledgments}
We thank the staffs at Fermilab and collaborating institutions,
and acknowledge support from the
DOE and NSF (USA);
CEA and CNRS/IN2P3 (France);
FASI, Rosatom and RFBR (Russia);
CNPq, FAPERJ, FAPESP and FUNDUNESP (Brazil);
DAE and DST (India);
Colciencias (Colombia);
CONACyT (Mexico);
KRF and KOSEF (Korea);
CONICET and UBACyT (Argentina);
FOM (The Netherlands);
STFC and the Royal Society (United Kingdom);
MSMT and GACR (Czech Republic);
CRC Program and NSERC (Canada);
BMBF and DFG (Germany);
SFI (Ireland);
The Swedish Research Council (Sweden);
and
CAS and CNSF (China).
\end{acknowledgments}

\bigskip 
\bibliographystyle{h-physrev3}
\bibliography{d0LQ1genRef}

%
%
%

%
%

\end{document}